\documentclass[12pt,journal,onecolumn]{IEEEtran}
\usepackage{algorithm}
\usepackage{algpseudocode}
\usepackage{subfigure}
\usepackage{epsfig}
\usepackage{graphicx}
\usepackage{float}
\usepackage{psfrag}
\usepackage{amsfonts,amsmath,amssymb}

\newtheorem{theorem}{Theorem}

\newtheorem{corollary}[theorem]{Corollary}

\newenvironment{proof}[1][Proof]{\begin{trivlist}
\item[\hskip \labelsep {\bfseries #1}]}{\end{trivlist}}

\newcommand{\qed}{\nobreak \ifvmode \relax \else
      \ifdim\lastskip<1.5em \hskip-\lastskip
      \hskip1.5em plus0em minus0.5em \fi \nobreak
      \vrule height0.75em width0.5em depth0.25em\fi}
            
\begin{document}
\title{Interference Alignment with Partially Coordinated Transmit Precoding}
\author{Aimal~Khan~Yousafzai~and~Mohammad~Reza~Nakhai}
\author{\normalsize{Aimal~Khan~Yousafzai,~\IEEEmembership{\normalsize{Student~Member,}}~and~Mohammad~Reza~Nakhai,~\IEEEmembership{\normalsize{Senior~Member,~IEEE}}}
\thanks{The authors are with the Center for Telecommunications Research, King's College London, Strand, WC2R~2LS, UK. (e-mail: $\left\{\textrm{aimal.yousafzai,reza.nakhai}\right\}$@kcl.ac.uk). }%
}
\maketitle
%
%

\begin{abstract}
\textbf{
In this paper, we introduce an efficient interference alignment (IA) algorithm exploiting partially coordinated transmit precoding to improve the number of concurrent interference-free transmissions, i.e., the multiplexing gain, in multicell downlink. The proposed coordination model is such that each base-station simultaneously transmits to two users and each user is served by two base-stations. First, we show in a $K$-user system operating at the information theoretic upper bound of degrees of freedom (DOF), the generic IA is proper when $K \leq 3$, whereas the proposed partially coordinated IA is proper when $K \leq 5$. Then, we derive a non-iterative, i.e., one shot, IA algorithm for the proposed scheme when $K \leq 5$. We show that for a given latency, the backhaul data rate requirement of the proposed method grows linearly with $K$. Monte-Carlo simulation results show that the proposed one-shot algorithm offers higher system throughput than the iterative IA at practical SNR levels.
}
\end{abstract}
%
%
\section{Introduction}
The problem of inter-cell interference in cellular networks has been the major impediment in delivering a consistent level of quality of service to geographically distributed users with an efficiently high spectral efficiency in terms of bits/sec/Hz/user. In a recent attempt to break the interference barrier, the information theoretic concept of interference alignment (IA) was used. This concept was initially introduced by Maddah-Ali \textit{et. al} to solve the MIMO $X$-channel \cite{maddah-ali}. In \cite{jafar1}, authors applied the concept of IA to $K$-user interference channel (IC) and obtained the theoretical upper bound of multiplexing gain or degrees of freedom (DOF). However, the practical development of IA in cellular networks is highly complex, e.g., an iterative process has been suggested in \cite{jafar2} to design the transmit and receive beamformers. 
In this paper, we modify the abstract distributed IA \cite{jafar2}, by introducing a level of structured coordination amongst the transmitters, i.e., base-stations (BSs), in the cellular network, to improve the achievable spectral efficiency at the cost of a limited overhead of backhaul processing. 
In our proposed coordination model, each user receives its desired signal from its dedicated BS, as the Primary transmitter, and a coordinating BS, as the secondary transmitter. In this chain of coordinated transmissions the first BS in the network coordinates with the second BS, the second BS coordinates with the third BS and so on, and eventually, the last BS coordinates with the first BS to complete the coordination cycle. In a sense, this transmission model is similar to Wyner circular model \cite{wyner},\cite{somekh2} as each user is supported by two BSs, but slightly departs from this model as each user is modeled to receive interference from all the BSs in the network. The coordination link between two BSs is uni-directional, e.g., the first BS shares its channel state information (CSI) with the second BS and serves as the secondary transmitter to the user of the second BS. As a result, each BS transmits to two users, i.e., to its own user as the primary transmitter and to the user of the coordinated BS as the secondary transmitter. 
\\
Through feasibility analysis, we compare the proposed coordination scheme with the generic IA. We establish some sufficient conditions for obtaining a proper system that can attain a feasible IA solution according to \cite{jafar3}. We show that, operating at half of the total available DOF, i.e., the information theoretic upper bound on DOF, the generic IA is proper for up to 3-user system, whereas the proposed partially coordinated scheme is proper for up to 5-user system for the same configuration, i.e., with the same number of transmit/receive antennas and the same number of DOF. Then, exploiting the rank deficiency of the interference covariance matrix of the reciprocal uplink channel, we propose an algorithm that implements the IA solution in the proposed coordination model in one shot. We establish a DOF upper bound for obtaining the one shot, i.e., non-iterative, solution and analyze the system configurations operating below the one shot upper bound. Furthermore, we analyze the backhaul overhead requirement of the proposed coordination model and show that the backhaul overhead in our model increases linearly with increase in the number of participating BSs. Simulation results confirm that our proposed scheme improves the sum rate performance compared to the iterative solution in \cite{jafar2}, for limited usage of the backhaul. 
\\
This paper is organized as follows. In section II, we introduce and formulate the proposed coordination scheme. In section III, we analyze and compare the feasibility of our proposed scheme with the generic IA and introduce a one shot IA solution for the proposed coordination scheme. Moreover, we compare the backhaul overhead in our proposed partially coordinated system with systems having fully coordinating BSs in terms of backhaul data rate requirement. In section IV, we support our analysis with simulation results and discussions. Finally, section V concludes this paper.
\\
The following notations are used throughout this paper. Bold lowercase and uppercase letters, e.g., $\mathbf{x}$ and $\mathbf{X}$, represent vectors and matrices, respectively. The $i^{th}$ dimension of $\mathbf{x}$ is shown by $x_i$ and $x_{ij}$ indicates the element at row $i$ and column $j$, i.e., $(i,j)^{th}$ element, in $\mathbf{X}$. Similarly, after partitioning a matrix into submatrices, we use calligraphic letter $\mathcal{X}_{ij}$ to define the submatrix at $\left(i,j\right)^{th}$ element of $\mathbf{X}$. 
The notation $\text{rem}(a,b)$ and $\text{mod}(\cdot)_{c}$ respectively represent the the remainder of the fraction $a/b$ and the modulo-arithmetic operation over $c$, where $a$, $b$ and $c$ are all integers. We use the operators $\lfloor \cdot \rfloor$, $\lceil \cdot \rceil$, $\text{rank} \left( \cdot \right)$, $\text{Null} \left( \cdot \right)$, $\text{Trace} \left( \cdot \right)$ and $\langle \cdot , \cdot \rangle$ to compute the floor, ceil, rank, null space, trace and dot product operations of their respective arguments. 
The operator $\text{Diag} (\cdot)$ generates a block-diagonal matrix using the elements of its argument. 
The superscripts $^T$, $^*$ and $^H$ represent the transpose, complex conjugate and conjugate transpose operations, respectively.
%
%
\section{System Description and Formulation}\label{sec2:sys_desc}
We consider a partially coordinated downlink transmission in a MIMO cellular system with $K$ BSs. In our proposed coordination, BSs are connected in pairs and in succession via backhaul links, as shown in Fig. \ref{fig:sysmodel}. We assume that the backhaul network is based on a line or a ring topology with the $k^{th}$ BS, $k=1,2,\ldots,K$, as the $k^{th}$ node in the line or the ring. The $k^{th}$ BS has $n_k$ transmit antennas and its corresponding user terminal has $m_k$ receiving antennas. We also assume time-division duplex (TDD) communication with synchronized time slots to ensure channel reciprocity and to use uplink CSI in the downlink. The successive pair-wise coordination among BSs is such that the $k^{th}$ BS establishes a coordination link with BS $k'$, where $k' = \text{mod}(k-2)_{K}+1$, 
in two steps as follows. First, BS $k'$ shares its CSI with BS $k$. Next, BS $k$ exploits this CSI to compute its transmit beamformer $\mathbf{W}_k \in \mathbb{C}^{\bar{n}_k \times d_k}$, where $\bar{n}_k=n_k+n_{k'}$ and $d_k$ is the DOF of user $k$. Then, $\mathbf{W}_k$ is used to transmit $d_k$ independent data streams towards user $k$ through the transmit antennas of both BSs $k$ and $k'$. In this coordination setup, the BSs $k$ and $k'$ serve as the primary and the secondary transmitters for user $k$, respectively. We assume flat fading channel that can be achieved by employing MIMO orthogonal frequency division multiplexing (OFDM) \cite{nakhai} as the baseband modulation. Moreover, we assume no intracell interference within a cell, as this can be obtained by using orthogonal frequency division multiple access (OFDMA). If the coordination among BSs is ignored then this system reduces to the standard $K$-user IC with $m_k \times n_k$ MIMO links. In the proposed scheme the overall received signal $\mathbf{y}= [\mathbf{y}^T_1 \ \mathbf{y}^T_2 \ \ldots \ \mathbf{y}^T_K]^T$, with $\mathbf{y}_k \in \mathbb{C}^{d_k \times 1}$ as the $k^{th}$ user's received signal, can be expressed as
\begin{equation}\label{eqn3:io}
\mathbf{y} = \mathbf{U}^H \big(\mathbf{H} \mathbf{V} \mathbf{x + n} \big) =\mathbf{U}^H \mathbf{H} \mathbf{V} \mathbf{x}  +  \mathbf{U}^H \mathbf{n}, 
\end{equation}
where $\mathbf{x}= [\mathbf{x}^T_1 \ \mathbf{x}^T_2 \ \ldots \ \mathbf{x}^T_K]^T$ and $\mathbf{n}= [\mathbf{n}^T_1 \ \mathbf{n}^T_2 \ \ldots \ \mathbf{n}^T_K]^T$, with $\mathbf{x}_k \in \mathbb{C}^{d_k \times 1}$ and $\mathbf{n}_k \in \mathbb{C}^{m_k \times 1}$, respectively, representing the transmitted symbol vector and the noise vector for user $k$. Furthermore, $\mathbf{H} \in \mathbb{C}^{\hat{m} \times \hat{n}}$, with $\hat{m}=\sum_{k=1}^{K}{m_k}$ and $\hat{n}=\sum_{k=1}^{K}{n_k}$, is the overall channel matrix which is composed of $K^2$ submatrices $\mathcal{H}_{i,j} \in \mathbb{C}^{m_i \times n_j}$, $i,j=1,2,\ldots,K$, as the $(i,j)^{th}$ element of $\mathbf{H}$. $\mathcal{H}_{i,j}$ represents the $m_i \times n_j$ MIMO channel between BS $j$ and user $i$. In \eqref{eqn3:io}, the overall receiver beamformer matrix is indicated by a block-diagonal matrix $\mathbf{U}=\text{Diag}(\mathbf{U}_1, \mathbf{U}_2, \ldots, \mathbf{U}_K)$, where the columns of $\mathbf{U}_k \in \mathbb{C}^{m_k \times d_k}$ form the receiver beamforming vectors of user $k$. The overall transmit beamforming matrix $\mathbf{V} \in \mathbb{C}^{\hat{n} \times \hat{d}}$ in \eqref{eqn3:io}, with $\hat{d} = \sum_{k=1}^{K}{d_k}$, is given by
\begin{equation}\nonumber 
\mathbf{V}=
\left[\begin{array}{c c c c c}
\mathbf{V}_1					&		\mathbf{\tilde{V}}_2	&		0					& \ldots										& 		0											\\
			0								&		\mathbf{V}_2					&		\ddots		&	\ddots										& 	\vdots									\\
			\vdots					& 			0									&		\ddots		&	\mathbf{\tilde{V}}_{K-1}	& 	0												\\
			0								&		\vdots								&		\ddots		& \mathbf{V}_{K-1}					&		\mathbf{\tilde{V}}_{K}	\\
\mathbf{\tilde{V}}_1	& 	0											& 	\ldots		& 		0											& 	\mathbf{V}_K
\end{array}\right],
\end{equation}
where the columns of $\mathbf{V}_k \in \mathbb{C}^{n_k \times d_k}$ and $\mathbf{\tilde{V}}_k \in \mathbb{C}^{n_{k'} \times d_k}$ are the distributed transmit beamforming vectors towards user $k$ through the antennas of BS $k$ and BS $k'$, respectively. Hence, the received vector by user $k$ can be written as
\begin{align}
\mathbf{y}_k 	={}& \mathbf{U}^H_k \Big( \overbrace{ \big( \mathcal{H}_{kk}\mathbf{V}_k + \mathcal{H}_{kk'}\mathbf{\tilde{V}}_k \big) \mathbf{x}_k}^{\text{Desired Signal}} \nonumber \\
							{}&	+ \underbrace{ \sum_{i=1, i\neq k}^{K}{\big( \mathcal{H}_{ki}\mathbf{V}_i + \mathcal{H}_{ki'}\mathbf{\tilde{V}}_i \big) \mathbf{x}_i} }_{\text{Interference Signal}} + \mathbf{n}_k \Big), \quad k=1,2,\ldots,K \label{eqn3:io_userk}, 
\end{align}
where $i'=\text{mod}(i-2)_{K}+1$. 
The solution to \eqref{eqn3:io} or \eqref{eqn3:io_userk} is attained by selecting $\mathbf{U}_k$, $\mathbf{V}_k$ and $\mathbf{\tilde{V}}_k$ such that,
\begin{align}
\mathbf{U}^H_k \big(\mathcal{H}_{ki}\mathbf{V}_i + \mathcal{H}_{ki'}\mathbf{\tilde{V}}_i \big) &= \mathbf{0} \qquad \forall \ k \neq i \ \text{and} \nonumber \\
\quad \text{rank}\Big(\mathbf{U}^H_k\big(\mathcal{H}_{kk}\mathbf{V}_k + \mathcal{H}_{kk'}\mathbf{\tilde{V}}_k \big)\Big) &= d_k \qquad \text{for} \ k = 1,2,\ldots,K \label{eqn3:sol_req}
\end{align}
To solve for the transmit and the receive beamformers in \eqref{eqn3:sol_req}, we first transform \eqref{eqn3:io} to an equivalent interference channel problem as follows. 
\subsection{An Equivalent Interference Channel Model}
Let us define the permutation matrix $\mathbf{P} = \big[\mathbf{P}_1 \ \mathbf{P}_2 \ \ldots \ \mathbf{P}_K \big]$, where $\mathbf{P}_k = \big[ \mathbf{1}_k \ \ \mathbf{1}_{k'} \big]$ for $k=1$, and $\mathbf{P}_k = \big[ \mathbf{1}_{k'} \ \ \mathbf{1}_{k} \big]$ for $k=2,\ldots,K$ with $k'=\text{mod}(k-2)_{K}+1$ and $\mathbf{1}_i \in \mathbb{R}^{\hat{n} \times n_i}$ given as,
\begin{equation}\label{eqn3:one_k}
\mathbf{1}_i = \Big[\mathcal{O}^T_{1} \ldots \mathcal{O}^T_{i-1} \ \ \mathcal{I}^T_{i} \ \ \mathcal{O}^T_{i+1} \ldots \mathcal{O}^T_{K} \Big]^T, \quad i=1,2,\ldots,K.
\end{equation}
In \eqref{eqn3:one_k}, $\mathcal{I}_{i}$ is $n_i \times n_i$ identity matrix and $\mathcal{O}_{j}$, for $j=1,2,\ldots,K$ and $j \neq i$, represents the $n_j \times n_i$ matrix with all zero elements. We define an equivalent interference channel matrix $\mathbf{G}=\mathbf{H}\mathbf{P}$ which can be partitioned into $K^2$ submatrices $\mathcal{G}_{i,j} \in \mathbb{C}^{m_i \times \bar{n}_j}$, with $\bar{n}_j=n_j+n_{j'}$, and $i,j \in \{ 1,2,\ldots,K\} $, where, 
\begin{equation}\label{eqn3:gij}
\mathcal{G}_{i,j}=\left\{
\begin{array}{l l}
\left[\mathcal{H}_{i,1} \ \mathcal{H}_{i,K} \right],	&		j = 1 	\\
\left[\mathcal{H}_{i,j-1} \ \mathcal{H}_{i,j} \right],	&		j = 2, \ldots K, 	
\end{array} \right.
\end{equation}
represents the submatrix at the $(i,j)^{th}$ location in $\mathbf{G}$. Next, we define the block-diagonal matrix $\mathbf{W} = \text{Diag} (\mathbf{W}_1, \mathbf{W}_2, \ldots, \mathbf{W}_K)$, where its $j^{th}$ element $\mathbf{W}_j \in \mathbb{C}^{\bar{n}_j \times d_j}$ is given as,
\begin{equation}\label{eqn3:Wj}
\mathbf{W}_j=\left\{
\begin{array}{l l}
\big[\mathbf{V}^T_j 				\		\mathbf{\tilde{V}}^T_j	 		\big]^T ,	&		j = 1 	\\
\big[\mathbf{\tilde{V}}^T_j \		 \mathbf{V}^T_j  					\big]^T	, &		j = 2, \ldots K.
\end{array} \right.
\end{equation}
Using \eqref{eqn3:Wj}, it can be easily verified that $\mathbf{W}=\mathbf{P}^T\mathbf{V}$. Therefore, $ \mathbf{G} \mathbf{W} = \mathbf{H} \mathbf{P} \mathbf{P}^T \mathbf{V} = \mathbf{H} \mathbf{V} $, as the permutation matrix satisfies $\mathbf{P} \mathbf{P}^T = \mathbf{I}$. Using this relation, we transform our partially coordinated multicell model into a $K$-user $m_k \times \bar{n}_k$ MIMO IC problem, where $\mathbf{W}_k$ is the transmit beamformer for user $k$. Thus, \eqref{eqn3:io} and \eqref{eqn3:io_userk} can be rewritten as,
\begin{equation}\label{eqn3:io_IC}
\mathbf{y} = \mathbf{U}^H \mathbf{G} \mathbf{W} \mathbf{x}  +  \mathbf{U}^H \mathbf{n}, 
\end{equation}
and
\begin{equation}\label{eqn3:io_userk_IC}
\mathbf{y}_k = \mathbf{U}^H_k \Big( \mathcal{G}_{k,k}\mathbf{W}_k \mathbf{x}_k + \sum_{i=1,i\neq k}^{K}{\mathcal{G}_{k,i}\mathbf{W}_i \mathbf{x}_i} + \mathbf{n}_k \Big), 
\end{equation}
respectively. Similarly, according to the IA principles \cite{jafar1}, the design constraints in \eqref{eqn3:sol_req} become
\begin{align}
\mathbf{U}^H_k \mathcal{G}_{k,i}\mathbf{W}_i &= \mathbf{0}, \qquad \forall \ k \neq i \ \text{and} \nonumber \\
\quad \text{rank}\big(\mathbf{U}^H_k\mathcal{G}_{k,k}\mathbf{W}_k\big) &= d_k, \qquad \text{for} \ i,k = 1,2,\ldots,K. \label{eqn3:sol_req2}
\end{align}
\section{Interference Alignment with Partially Coordinated Precoding}
In this section, first, we analyze the sufficient conditions for obtaining a proper uncoordinated generic interference channel system in terms of maximum number of users supported at the information theoretic upper bound on DOF and compare it with our proposed partial coordination scheme, in \eqref{eqn3:sol_req2}.
\subsection{Feasibility Analysis}\label{subsec:feasibility}
According to \cite{jafar3}, a generic $K$-user IC with $m_k \times n_k$ MIMO links, denoted as a $\prod_{k=1}^{K}{(m_k \times n_k,d_k)}$, is a proper system if the number of variables involved in each and every subset of equations is greater than or equal to the number of equations in that subset. Based on the intuitive discussion in \cite{jafar3} (see sec. V and VI), authors argue that a proper IC system is likely to attain a feasible IA solution. In this context, a symmetric IC system, with $m_k=m$, $n_k=n$ and $d_k=d$, $\forall \ k$, described as $(m \times n, d)^K$, is proper if the total number of equations is less than or equal to the total number of variables, i.e., $m + n \geq (K+1)d$. 
Moreover, in \cite{jafar3}, the various groups of IC systems are classified such that within a group $m + n = c$, where $c$ is a constant, and $\text{min}(m ,n ) \geq d $, $\forall \ k$. Based on the classification of a group of systems, the authors in \cite{jafar3} establish a rule of thumb which states that if any system within a group is proper then almost all the systems of that group are also proper and vice versa. 
\\
Following the definitions in \cite{jafar3}, our equivalent IC system in \eqref{eqn3:io_IC} can be modeled as $\prod_{k=1}^{K}{(m_k \times \bar{n}_k,d_k)}$. Furthermore, the information theoretic upper bound for DOF, introduced in \cite{jafar1}, can be tightly approximated as $\hat{d} =  \sum_{k=1}^{K}{(m_k+n_k-\text{rem}(m_k+n_k,2))/4}$ for the general $K$-user IC and $ \hat{d} = \lfloor \frac{\sum_{k=1}^{K}{\lfloor  \frac{m+n}{2} \rfloor}}{2}\rfloor = \lfloor K (c-\text{rem}(c,2))/4 \rfloor$, for the symmetric $K$-user case, where $\text{rem}(c,2)$ is 0 or 1 for an even or odd values of $c$, respectively. Note that $\hat{d}$ is not necessarily an integer multiple of $K$. This requires to privilege a number of users by allowing them to operate at a higher integer number of DOF, i.e., $\lceil \hat{d}/K \rceil$, than the others, i.e., operating at $\lfloor \hat{d}/K \rfloor$. However, this may not be the best policy of allocating the DOF resources amongst the users due to the fairness considerations as well as the fact that different users experience different fading conditions in the channel. Clearly, assigning a larger DOF to a user facing an aggressive fading condition in the channel is not an efficient policy, as it may require a higher transmit power to deliver a prescribed level of data rate per stream. In the following, we propose a time-sharing strategy over a number of time-slots so that the available DOF resources are equally distributed, when averaged over all time slots, amongst all $K$ users.
\begin{theorem}\emph{(A uniformly time-shared distribution of DOF in the generic IA)}\label{theorem1}
A system of $K$-user asymmetric interference channel described by $\prod_{k=1}^{K}{(m \times n,d_k)}$, where user $k$ is operating at an integer number of $d_k$ of DOF so that $\hat{d} =\sum_{k=1}^{K}{d_k} = \lfloor K (c-\text{rem}(c,2))/4 \rfloor$ is the information theoretic upper bound, can be described as an equivalent symmetric system of $(m \times n, \hat{d}/K)^K$ defined under a time sharing schedule over a span of ${\binom{K}{\alpha}}$ time slots, where $\alpha = \mathrm{rem}(\hat{d},K)$.
\end{theorem}
\begin{proof}
Let $\tau = {\binom{K}{\alpha}}$ be the total number of allocated time slots, $\alpha$ be the remaining number of DOF to be shared uniformly among $K$ users, and $\theta = \frac{\alpha}{K} \tau = {\binom{K - 1}{\alpha - 1}}$. In the equivalent symmetric system each user $k$ operates at a DOF of $\lceil \frac{\hat{d}}{K} \rceil$ for the duration of $\theta$ time slots and at a DOF of $ \lfloor \frac{\hat{d}}{K} \rfloor$ for the remaining $\tau - \theta$ time slots. The distribution of DOF is attained such that in each of the $\tau$ time slots the total number of independent data streams in the interference channel system is $\hat{d}$. Over the span of $\tau$ time slots, each user $k$ attains a total of $\theta \lceil \frac{\hat{d}}{K} \rceil + (\tau - \theta) \lfloor \frac{\hat{d}}{K} \rfloor = \tau \lfloor \frac{\hat{d}}{K} \rfloor + \theta $ DOF. Effectively, the DOF per time slot for each user $k$, i.e., $d_k$, is given as
\begin{align}
d_k	&=	\frac{\tau \lfloor \frac{\hat{d}}{K} \rfloor + \theta}{ \tau} = \lfloor \frac{\hat{d}}{K} \rfloor + \frac{\alpha}{K}  \nonumber	\\
		&= \frac{\hat{d}-\alpha}{K} + \frac{\alpha}{K} = \frac{\hat{d}}{K}. \label{th0}
\end{align}
Thus, over the span of $\tau$ time slots the system can be modeled as $(m \times n, \hat{d}/K)^K$ system.
\end{proof}

\textit{Example 1:} 
Consider an asymmetric 5-user interference channel with $3 \times 3$ MIMO links where each user operates at its information theoretic upper bound of DOF, i.e., $\hat{d} = \lfloor 5 \times 6 /4 \rfloor = 7$. This system can be described as $(3 \times 3, 2)^2(3 \times 3, 1)^3$. Applying Theorem \ref{theorem1}, one can easily verify that after assigning one DOF per user the remaining number of DOF to be allocated is $\alpha = 2$, the total number of required time slots is $\tau = \binom{5}{2}=10$ and $\theta = (10) 2/5 = 4$. Hence, over the duration of 10 time slots, each user operates at $\lceil 7 / 5 \rceil = 2$ DOF for 4 time slots and at $\lfloor 7 / 5 \rfloor = 1$ DOF for the remaining 6 time slots, i.e., a total number of 14 DOF. Therefore over a span of 10 time slots, the equivalent symmetric system  can be modeled as $(3 \times 3, 7/5)^5$, where each user attains $7/5$ DOF per time slot. 
\\
In the sequel, we calculate the upper bounds on the number of users that can be served at the information theoretic limit of DOF in the generic and the proposed partially coordinated proper systems.
\begin{corollary}\emph{(An upper bound on $K$ for generic IA)}\label{corollary1}
A system of $K$-user interference channel described by $\prod_{k=1}^{K}{(m \times n ,d_k)}$ that operates at the information theoretic upper bound $\hat{d} = \lfloor \frac{K}{4}(c - \text{rem}(c,2))\rfloor$, where $c=m+n$, $\text{min}(m,n) \geq \lceil \hat{d}/K \rceil$, and $\text{rem}(c,2)$ is either 0 or 1 for even or odd values of $c$, respectively, is proper if 
\begin{equation}\label{th1}
K \leq 3 + 4 \frac{\text{rem}(c,2)}{c-\text{rem}(c,2)}.
\end{equation}
\end{corollary}
\begin{proof}
An asymmetric system $\prod_{k=1}^{K}{(m \times n,d_k)}$ can be modeled as an equivalent symmetric system $(m \times n , \tilde{d})^K$, where $\tilde{d}= \hat{d}/K \approx \frac{1}{4}(c -\text{rem}(c,2))$, using the time sharing concept of DOF, developed in Theorem \ref{theorem1}. To obtain a proper symmetric system, we require the total number of equations $N_e$ and the total number of variables $N_v$ satisfy $N_e \leq N_v$, i.e.,
\begin{align}
K \tilde{d}													&{}\leq c - \tilde{d}				\nonumber \\
K\big( c - \text{rem}(c,2) \big)		&{}\leq 3c + \text{rem}(c,2) 	\nonumber \\
K																		&{}\leq 3 + 4\frac{\text{rem}(c,2)}{c - \text{rem}(c,2)} \nonumber
\end{align}
which is a sufficient condition for obtaining a proper system.
\end{proof}

\textit{Remark 1:}
It can be easily verified from \eqref{th1} that the system $\prod_{k=1}^{K}{(m \times n ,d_k)}$ is proper if $K \leq 4$. The equality of $K=4$ holds only for $c = 5$ and for $c \neq 5$, we have $K \leq 3$. The exceptional case of $K=4$ with $c=5$ applies to the group of symmetric systems $(1 \times 4, 1)^4$, $(2 \times 3, 1)^4$, $(3 \times 2, 1)^4$, $(4 \times 1, 1)^4$ systems that are given as Examples 1, 3 and 5 in \cite{jafar3}. Furthermore, generic IC systems operating at their information theoretic upper bound of DOF with $K \geq 4$, i.e., with an exception of $c=5$ for $K=4$, are not proper and, hence, are not feasible to attain the IA solution. 
This fact is confirmed in Fig. 4 of \cite{jafar2}, where the DOF upper bound cannot be achieved for four users with 4 and 5 antenna cases. We will also confirm this through simulation results presented in section \ref{sec_sim}. \\
Next, we establish an upper bound on $K$ for a proper partially coordinated IA system. 
\begin{corollary}\emph{(An upper bound on $K$ for IA with partial coordination)}\label{corollary2}
A system of $K$-user partially coordinated channel described by the equivalent interference channel representation $\prod_{k=1}^{K}{(m \times 2n ,d_k)}$, that operates at the information theoretic upper bound $\hat{d} = \lfloor \frac{K}{4} (c - \text{rem}(c,2)) \rfloor$, where $c = m + n$, $\text{min}(m,n) \geq \lceil \hat{d}/K \rceil$ and $\text{rem}(c,2)$ is either 0 or 1 for even or odd values of $c$, respectively, is proper if 
\begin{equation}\label{th2}
K \leq 3 + 4\frac{n + \text{rem}(c,2)}{m + n - \text{rem}(c,2)}. 
\end{equation}
\end{corollary}
\begin{proof}
An asymmetric system $\prod_{k=1}^{K}{(m \times 2n,d_k)}$ can be modeled as an equivalent symmetric system $(m \times 2n , \tilde{d})^K$, where $\tilde{d}= \hat{d}/K \approx \frac{1}{4}(c -\text{rem}(c,2))$, using the time sharing concept of DOF, developed in Theorem \ref{theorem1}. One can easily verify that, here, the total number of equations $N_e = K\tilde{d}^2(K-1)$ and the total number of variables $N_v = \sum_{k=1}^{K}{\tilde{d}(m + 2n  -2\tilde{d})} = \tilde{d} (\sum_{k=1}^{K}{(m + n )}+\sum_{k=1}^{K}{n}-2K\tilde{d}) = 
K\tilde{d}(c + n -2\tilde{d})$. To obtain a proper system, we require $N_e \leq N_v$, i.e.,
\begin{align}
K \tilde{d}													&{}\leq c + n - \tilde{d}				\nonumber \\
K																		&{}\leq 3 + 4\frac{ n + \text{rem}(c,2)}{c - \text{rem}(c,2)} \nonumber \\
																		&{}\leq 3 + 4\frac{ n  + \text{rem}(c,2)}{ m  + n  - \text{rem}(c,2)} \nonumber
\end{align}
which is a sufficient condition for obtaining a proper system of IA with partial coordination.
\end{proof}

\textit{Remark 2:} It can be easily verified from \eqref{th2} that the equivalent IC system $\prod_{k=1}^{K}{(m \times 2n ,d_k)}$ is proper if $K \leq 5$, when $m = n  \pm \text{rem}(c,2)$, $\forall \ k$, and its information theoretic upper bounds on DOF for $K=3$, $K=4$ and $K=5$ are given as $\hat{d} = \lfloor\frac{3}{4}(c - \text{rem}(c,2)) \rfloor = \lfloor 3 n/2 \rfloor $, $\hat{d} = \lfloor (c - \text{rem}(c,2)) \rfloor = 2n$ and $\hat{d} = \lfloor\frac{5}{4}(c - \text{rem}(c,2)) \rfloor = \lfloor 5 n/ 2 \rfloor$, respectively. \\
In the sequel, we introduce a one-shot algorithm that achieves DOF upper limits within $\hat{d}\leq2n$ by IA with partial coordination and discuss, in more details, the achievability of various DOF configurations for $K=3$, $K=4$ and $K=5$.    
%
\subsection{A one shot solution for IA with Partial Coordination Scheme}\label{subsec:oneshotsol}
Here, we describe a one shot IA solution to find the receive and the transmit beamformers, i.e., $\mathbf{U}_i \in \mathbb{C}^{m_i \times d_i}$, $i=1,2,\ldots,K$, and $\mathbf{W}_k \in \mathbb{C}^{\bar{n}_k \times d_k}$, $k=1,2,\ldots,K$, respectively, for the proposed partial coordination scheme in two steps. First, we design $\mathbf{U}_i$, $\forall \ i$, for any $i^{th}$ receive beamformer such that the signal power received by each user $i$ in the original network is maximized. Then, using these receive beamformers in the reciprocal network, where roles of the transmitters and the receivers are reversed, we design $\mathbf{W}_k$, $\forall \ k$, such that the interference leakage at the receiving ends of the reciprocal network is minimized to zero. These steps are described as follows. Note that in the following analysis, we use the notation defined in \cite{jafar2} to demonstrate the variables in the reciprocal network, i.e., in the reverse link, with a left arrow on top.

\underline{\textit{Step 1}}: In the original network, the desired signal power at receiver $i$ due to the partially coordinated transmission from BSs $i$ and $i'$ is given by
\begin{equation}\label{eqn:s_i}
S_i = \mathrm{Trace} \left( \mathbf{U}^H_{i} \mathbf{Q}_i \mathbf{U}_i \right), 
\end{equation}
where
\begin{equation}\nonumber
\mathbf{Q}_i = \frac{P_i}{d_i} \mathbf{\cal{G}}_{i,i} \mathbf{W}_{i} \mathbf{W}^H_{i} \mathbf{\cal{G}}_{i,i}^H
\end{equation}
is the signal covariance matrix at user $i$ and $P_i$ is the power of message signal $\mathbf{x}_i$ transmitted through BSs $i$ and $i'$, i.e., the $i^{th}$ transmitter in the equivalent IC model. First of all, the $i^{th}$ transmitter computes the singular value decomposition of $\mathcal{G}_{i,i} \in \mathbb{C}^{m_i \times \bar{n}_i}$, $i=1,2,\ldots,K$, as $\mathcal{G}_{i,i} = \mathbf{F}_i \mathbf{\Sigma}_i \mathbf{M}^H_i$, where $\mathbf{F}_i=[\mathbf{f}^{[i]}_{1} \ldots \mathbf{f}^{[i]}_{m_k}]$, $\mathbf{\Sigma}_i=\text{Diag}(\sigma^{[i]}_{1} \ldots \sigma^{[i]}_{m_k})$ and $\mathbf{M}_i=[\mathbf{m}^{[i]}_{1} \ldots \mathbf{m}^{[i]}_{m_k}]$, and initializes $\mathbf{W}_i = \tilde{\mathbf{M}}_i = [\mathbf{m}^{[i]}_{1} \ldots \mathbf{m}^{[i]}_{d_i}]$, i.e., the first $d_i$ columns of $\mathbf{M}_i$. Then, the receiver $i$ solves the optimization problem 
$\displaystyle {\max_ {\mathbf{U}_i : m_i \times d_i, \ \mathbf{U}_i\mathbf{U}_i^H= \mathbf{I}_{d_i}} S_i}$ 
to find the receive beamformers $\mathbf{U}_i$ that maximizes $S_i$ in \eqref{eqn:s_i}. It can be easily verified that \eqref{eqn:s_i} is maximized when $\mathbf{U}_i=[\mathbf{f}^{[i]}_{1} \ldots \mathbf{f}^{[i]}_{d_i}]$, i.e., the first $d_i$ columns of $\mathbf{F}_i$, and the maximum received signal power $S^{\mathrm{max}}_i$ is calculated as
\begin{equation}\label{eqn:s_i1}
S^{\mathrm{max}}_i=\frac{P_i}{d_i}\sum_{j=1}^{d_i}{(\sigma^{[i]}_{j})^2}
\end{equation}
which is equivalent to the received signal power at user $i$ with $d_i$ DOF in the single user scenario.

\underline{\textit{Step 2}}: 
In the reciprocal network, where the $k^{th}$ user is the $k^{th}$ transmitter, i.e., $\overleftarrow{\mathbf{W}}_k = \mathbf{U}_k$, and the coordinating BSs $k$ and $k'$ form the $k^{th}$ receiver, i.e., $\overleftarrow{\mathbf{U}}_k = \mathbf{W}_k$, the total interference leakage at the $k^{th}$ receiver due to all undesired transmitters $l, l\neq k,$ is given by
\begin{equation}\nonumber
\overleftarrow{I}_{k} = \mathrm{Trace} \left( \overleftarrow{\mathbf{U}}^H_{k} \overleftarrow{\mathbf{Q}}_k \overleftarrow{\mathbf{U}}_k \right) = \mathrm{Trace} \left( \mathbf{W}^H_{k} \overleftarrow{\mathbf{Q}}_k \mathbf{W}_k \right),
\end{equation} 
where
\begin{align}\nonumber
\overleftarrow{\mathbf{Q}}_{k} 	&= \sum_{l=1,l \neq k}^{K} {\frac{\overleftarrow{P}_l}{d_l} \overleftarrow{\mathbf{\cal{G}}}_{k,l} \overleftarrow{\mathbf{W}}_l \overleftarrow{\mathbf{W}}^H_l \overleftarrow{\mathbf{\cal{G}}}_{k,l}^H} \nonumber \\
																&= \sum_{l=1,l \neq k}^{K} {\frac{\overleftarrow{P}_l}{d_l} \mathbf{\cal{G}}^H_{l,k} \mathbf{U}_l \mathbf{U}^H_l  \mathbf{\cal{G}}_{l,k}} \nonumber
\end{align}
is the interference covariance matrix at receiver $k$, $\overleftarrow{P}_l$ is the power of the message signal $\overleftarrow {\mathbf{x}}_l$ transmitted in the reverse link from user $l$ and due to channel reciprocity $\overleftarrow{\mathbf{\cal{G}}}_{k,l} = \mathbf{\cal{G}}^H_{l,k}$, $\forall \ l,k$. The $k^{th}$ receiver in the reverse link solves the optimization problem 
$\displaystyle{\min_{ \mathbf{W}_{k} : \bar{n}_k \times d_k, \ \mathbf{W}^H_{k} \mathbf{W}_{k} = \mathbf{I}_{d_k}} \overleftarrow{I}_{k}}$ 
to find $\mathbf{W}_{k}$ that minimizes the interference leakage due to all undesired transmissions, i.e., all users except user $k$. This optimization problem designs $\mathbf{W}_{k}$ to be the eigenvectors corresponding to $d_k$ smallest eigenvalues of $\overleftarrow{\mathbf{Q}}_{k}$. Note that the interference covariance matrix at the $k^{th}$ receiver in the reciprocal channel, i.e., $\overleftarrow{\mathbf{Q}}_k \in \mathbb{C}^{\bar{n}_k \times \bar{n}_k}$ has a rank of $r_k = \sum_{i=1,i\neq k}^{K}{d_i} = \hat{d}-d_k$. By defining $a_k = \textrm{Nullity} (\overleftarrow{\mathbf{Q}}_k) = \bar{n}_k - r_k $, where $\textrm{Nullity} (\overleftarrow{\mathbf{Q}}_k)$ is the number of dimensions in the null-space of $\overleftarrow{\mathbf{Q}}_k$ \cite{golub}, one can easily verify that $d_k \leq a_k$, $\forall \ k$, when $\hat{d}\leq \bar{n}$, where $\bar{n}= \text{min}(\bar{n}_1, \ldots, \bar{n}_K)$ for asymmetric systems and $\bar{n}= 2n$ for symmetric systems. Hence, $\mathbf{W}_k$ is chosen in the null-space of $\overleftarrow {\mathbf{Q}}_k$ to suppress the interference leakage $\overleftarrow{I}_k$ in the reciprocal network. 
\\
Next, we show that choosing $\mathbf{W}_k$ in the null-space of $\overleftarrow {\mathbf{Q}}_k$ and using its columns as the transmit beamforming vectors in the original network will suppress the interference to other unintended users $l\neq k$. Let us decompose $\overleftarrow {\mathbf{Q}}_k$ as $\overleftarrow{\mathbf{Q}}_k= \mathbf{R}_k\mathbf{R}^H_k$, where $\mathbf{R}_k = \tilde{\mathbf{G}}_k \tilde{\mathbf{U}}_k$, $\tilde{\mathbf{G}}_k = [\mathcal{G}^H_{1,k} \ldots \mathcal{G}^H_{k-1,k} \ \mathcal{G}^H_{k+1,k} \ldots \mathcal{G}^H_{K,k}]$ and $\tilde{\mathbf{U}}_k = \text{Diag}(\mathbf{U}_1 \ldots \mathbf{U}_{k-1} \ \mathbf{U}_{k+1} \ldots \mathbf{U}_{K})$. Since $\mathbf{W}^H_k\overleftarrow{\mathbf{Q}}_k\mathbf{W}_k= (\mathbf{R}^H_k \mathbf{W}_k )^H \mathbf{R}^H_k\mathbf{W}_k  = \mathbf{0}$, therefore, $\mathbf{R}^H_k \mathbf{W}_k = \mathbf{0}$. Finally, by expanding $\mathbf{R}_k$ in $\mathbf{R}^H_k \mathbf{W}_k = \mathbf{0}$, we arrive at $\mathbf{U}^H_l \mathcal{G}_{l,k}\mathbf{W}_k = \mathbf{0}$, $\forall \ l \neq k$, $l=1,2,\ldots,K$, which satisfies the first condition of the interference alignment solution in \eqref{eqn3:sol_req2}. Therefore, in this way, BSs $k$ and $k'$ can jointly transmit the message signal $\mathbf{x}_k$ to user $k$ with $d_k$ DOF without causing interference to the other users.
\\
Due to the update of $\mathbf{W}_i$, $i=1,2,\ldots,K$, in step 2, the received signal power at the $i^{th}$ user of the original network deviates from the single user performance, i.e., $S^{\mathrm{max}}_i$, in \eqref{eqn:s_i1}, as follows. Starting with \eqref{eqn:s_i} and substituting for $\mathbf{Q}_i$ with updated $\mathbf{W}_i$ values calculated in step 2 and $\mathbf{U}_i=\tilde{\mathbf{F}}_i$, we can write
\begin{align}
S_i & = \frac{P_i}{d_i} \ \mathrm{Trace}\left(\tilde{\mathbf{F}}_i^H \mathcal{G}_{i,i} \mathbf{W}_{i} \mathbf{W}^H_{i} \mathcal{G}_{i,i}^H\tilde{\mathbf{F}}_i \right) \nonumber\\
    & = \frac{P_i}{d_i} \ \mathrm{Trace}\left(\mathcal{G}_{i,i}^H \tilde{\mathbf{F}}_i \tilde{\mathbf{F}}_i^H \mathcal{G}_{i,i} \sum_{j=1}^{d_i} \mathbf{w}_j^{[i]} (\mathbf{w}_j^{[i]})^H \right)\nonumber
\end{align}
But, it can be easily verified that $\mathcal{G}_{i,i}^H \tilde{\mathbf{F}}_i\tilde{\mathbf{F}}_i^H \mathcal{G}_{i,i}=\tilde{\mathbf{M}}_i\tilde{\mathbf{\Sigma}}_i^2 \tilde{\mathbf{M}}_i^H$, therefore
\begin{align}
S_i & =\frac{P_i}{d_i} \sum_{j=1}^{d_i}{\left( (\mathbf{w}_j^{[i]})^H\tilde{\mathbf{M}}_i \tilde{\mathbf{\Sigma}}^2_i \tilde{\mathbf{M}}^H_i \mathbf{w}_j^{[i]} \right)}\nonumber\\
&= \frac{P_i}{d_i} \sum_{j=1}^{d_i}{(\sigma^{[i]}_{j})^2 \sum_{l=1}^{d_i}{|\langle \mathbf{m}^{[i]}_j , \mathbf{w}^{[i]}_l \rangle|^2}}, \label{eqn:s_i2}
\end{align}
where $\tilde{\mathbf{\Sigma}}_i=\text{Diag}(\sigma^{[i]}_1 \ldots \sigma^{[i]}_{d_i})$ and $\mathbf{W}_i=[\mathbf{w}^{[i]}_1 \ldots \mathbf{w}^{[i]}_{d_i}]$. A comparison of \eqref{eqn:s_i1} and \eqref{eqn:s_i2} shows that each term $(\sigma^{[i]}_{j})^2$, i.e., the channel gain corresponding to the $j^{th}$ largest singular values of $\mathcal{G}_{i,i}$, is degraded by a factor of $\sum_{l=1}^{d_i}{|\langle \mathbf{m}^{[i]}_j , \mathbf{w}^{[i]}_l \rangle|}^2$, where $\langle \mathbf{m}^{[i]}_j , \mathbf{w}^{[i]}_l \rangle$ is the projection of transmit beamformer $\mathbf{w}^{[i]}_l$, $l=1,\ldots,d_i$, on $\mathbf{m}^{[i]}_j$. Since $\mathbf{W}^H_i\mathbf{W}_i=\mathbf{I}_{d_i}$, $\forall \ i$, and $\tilde{\mathbf{M}}^H_i\tilde{\mathbf{M}}_i=\mathbf{I}_{d_i}$, $\forall \ i$, therefore, $|\langle \mathbf{m}^{[i]}_j , \mathbf{w}^{[i]}_l \rangle| \leq 1$. However, despite this degradation, the Monte-Carlo simulation results, described in section \ref{sec_sim}, show that the one shot partially coordinated IA offers higher sum rate than the distributed IA \cite{jafar2} in similar scenarios. 
\subsection{Achievable gains in IA with partial coordination}\label{subsec:gains}
In a symmetric distributed IA system \cite{jafar2}, with $c=m+n$ and $m=n+\text{rem}(c,2)$, $\forall \ k$, the achievable DOF information theoretic upper bound satisfies $\hat{d} = \lfloor K(c - \text{rem}(c,2))/4 \rfloor>n$ for $K>2$, whereas in partially coordinated IA, it is limited by $\hat{d} \leq 2n$, for $K \leq 4$, as established in \textit{Remark 2}. Furthermore, for a symmetric distributed IA, $a_k-d_k=n-\hat{d}<0$ or equivalently $a_k<d_k$, for $K>2$, because $\overleftarrow{\mathbf{Q}}_k$ has a rank of $r_k=\hat{d}-d_k$ and contains only $n$ columns. This implies that $\mathbf{W}_k$ cannot be designed in one shot in the null-space of $\overleftarrow{\mathbf{Q}}_k$ and an unpredictable number of iterations is needed to reduce the rank of $\overleftarrow{\mathbf{Q}}_k$. These iterations continue until the required number of dimensions of $d_k$ is reached for the null-space of $\overleftarrow{\mathbf{Q}}_k$. For instance, the distributed IA in a symmetric  interference channel system with $K=3$, $m=n=2$, $\hat{d}=3$, i.e., $d_k=1$, $\forall \ k$, requires iterative adjustment of transmit beamformers until $r_k=1$ is reached at each receiver. Whereas, for a symmetric IA system with partial coordination a DOF of $\hat{d}$ that satisfies $\hat{d}\leq 2n$ can be achieved in one shot because $d_k \leq a_k$, i.e, the null space of $\overleftarrow{\mathbf{Q}}_k$ has enough dimensions to contain $d_k$ DOF, or $d_k$ independent streams per user $k$, from the starting point of iterations. Let the columns of $\mathbf{T}_k \in \mathbb{C}^{2n \times a_k}$ form a set of orthonormal basis for the null-space of $\overleftarrow{\mathbf{Q}}_k$ and the columns of $\mathbf{W}_k  \in \mathbb{C}^{2n \times d_k}$ be selected from the columns of $\mathbf{T}_k$. Then, for each user $k$, we have the flexibility of choosing a set of $d_k$ columns from $E_k = {\binom{a_k}{d_k}}$ possible selections of sets for $\mathbf{W}_k$. Clearly, the optimum set would be a set of $d_k$ columns that maximizes the received signal power at user $k$ according to \eqref{eqn:s_i2}, i.e., $\displaystyle{\max_{ \mathbf{W}_k=[\mathbf{w}^{[k]}_1 \ldots \mathbf{w}^{[k]}_{d_k}]} S_k}$. As an alternative approach, one can set the columns of $\mathbf{W}_k$ such that the geometric mean of the individual SNR (see section 2.5 on p. 33 in \cite{cioffi_notes}) values of all the $d_k$ parallel independent channels of user $k$ is maximized.
\\
The steps of the one-shot IA algorithm to implement IC systems with the proposed partial coordination are summarized in Algorithm \ref{dist_coop_broadcast}, as follows. 
\begin{center}
\begin{algorithm}
\caption{One shot Algorithm for IA with Partially Coordinated Transmit Precoding}\label{dist_coop_broadcast}
\begin{algorithmic}[1]
\State First, design the receiver beamformers $\mathbf{U}_k$, $k=1,2,\ldots,K$. Set the $d^{th}$ vector of $\mathbf{U}_k$, $d=1,2,\ldots,d_k$, as $\mathbf{u}_{k,d} = \nu_d(\mathcal{G}_{k,k} \mathcal{G}^H_{k,k})$, where $\nu_d(\mathbf{X})$ represents the eigenvector that corresponds to the $d^{th}$ largest eigenvalue of $\mathbf{X}$. 
\State At each user $k$, using the CSI of BS $k'$, i.e., $\mathbf{H}_{k'}=[\mathcal{H}_{1,k'} \ \ldots \ \mathcal{H}_{K,k'}]$, construct $\mathbf{G}_{k}=[\mathcal{G}_{1,k} \ \ldots \ \mathcal{G}_{K,k}]$ according to \eqref{eqn3:gij}. Compute $\overleftarrow{\mathbf{Q}}_k$  and $\mathbf{T}_k = \text{Null}(\mathbf{Q}_k)$. 
\State For each user $k$, define $\mathbf{A}_i \in \mathbb{C}^{\bar{n}_k \times d_k}$, for $i=1,2,\ldots,E_K$, $E_k = {\binom{a_k}{d_k}}$, such that each matrix $\mathbf{A}_i$ contains a unique selection of $d_k$ columns of $\mathbf{T}_k$.
\State For each user $k$, compute $\mathbf{B}_i = \mathbf{U}_k \mathcal{G}_{k,k} \mathbf{A}_i$, $i=1,2,\ldots,E_K$, and find the eigenvalues of $\mathbf{B}_i$ as $\lambda^{[i]}_{1}, \ldots, \lambda^{[i]}_{d_k}$. Compute the geometric SNR coefficient $\gamma_i = \prod_{j=1}^{d_k}{|\lambda^{[i]}_{j}|}$, $\forall \ i$, and set $\mathbf{W}_k=\mathbf{A}_{\text{max}}$, where $\mathbf{A}_{\text{max}}$ is the matrix $\mathbf{A}_i$ corresponding to the largest value of $\gamma_i$.
\State Finally, set the primary and secondary transmit beamforming vectors $\mathbf{V}_k$ and $\tilde{\mathbf{V}}_k$, respectively, for each user $k$, according to \eqref{eqn3:Wj}.
\end{algorithmic}
\end{algorithm}
\end{center}

\textit{Definition 1:}
We call a system satisfying $\hat{d} < \bar{n}$, where $\bar{n} = \mathrm{min}(\bar{n}_1, \ldots, \bar{n}_K)$ in an asymmetric system and $\bar{n} = 2n$ in a symmetric system, or equivalently $d_k < a_k$, $\forall \ k$, as \textit{flexible} due to the flexibility of selecting $\mathbf{W}_k$ in this system. Similarly, a system with $\hat{d} = \bar{n}$ or equivalently $d_k = a_k$ , $\forall \ k$, is termed as \textit{rigid}. In a \textit{rigid} system $E_k = 1$ and therefore, $\mathbf{W}_k = \mathbf{T}_k$, $\forall \ k$. 
\\
In the following examples, we analyze various symmetric system configurations for $K=3$, $K=4$ and $K=5$ in the proposed partially coordinated scenario and design the transmit beamformers $\mathbf{W}_k$ such that the overall throughput is maximized. 

\textit{Example 2} ($K=3$): For $K=3$ symmetric systems, the information theoretic DOF upper limit is less than the number of columns of $\overleftarrow {\mathbf{Q}}_k$, $\forall \ k$, i.e., $\hat{d} = \lfloor 3 n /2 \rfloor < 2n$, and therefore, $d_k < a_k$, $\forall \ k$. Hence, partially coordinated IA can achieve this system with $\hat{d}$ DOF in one-shot. Furthermore, we can attain higher throughput for $K=3$ by increasing the number of independent data streams within $ \lfloor 3 n /2 \rfloor \leq \hat{d} \leq  2n$ interval. For example, with $m=n=2$, the proposed one shot solution achieves $\hat{d}=3$ and $\hat{d}=4$, whereas, the distributed IA algorithm \cite{jafar2} can only achieve $\hat{d}=3$ iteratively. Likewise, with $m=n=3$, the proposed partially coordinated IA achieves $\hat{d}=4$, $\hat{d}=5$ and $\hat{d}=6$ in one-shot, whereas, the distributed IA \cite{jafar2} can only attain $\hat{d}= 4$ through iterative process.

\textit{Example 3} ($K=4$): As established in \textit{Remark 1}, a conventional IC system with $K \geq 4$ is not proper \footnote{the exceptional feasible case for $K=4$ with $c=m+n=5$ is identified in \textit{Remark 1}.} at the information theoretic upper bound of DOF and, hence, cannot be achieved by the generic IA. This is confirmed in \cite{jafar2}, where the distributed IA achieves $K=4$ with $m=n=4$ at a total of $6$ degrees of freedom out of a theoretically possible $8$ and $K=4$ with $m=n=5$ at a total of $8$ degrees of freedom out of a theoretically possible $10$, (see Fig. 4 in \cite{jafar2}). Whereas, these theoretical upper bounds can be attained in one shot, as $\hat{d} \leq 2n$ in both cases, by introducing the proposed partial coordination into the interference channel. Furthermore, the iterative IA in \cite{jafar2} is unable to solve the conventional IC systems with $K=4, m=n=2$ at $\hat{d}=4$ and  $K=4, m=n=3$ at $\hat{d}=6$, whereas the introduced IA with partial coordination achieves both of these systems in one shot, because they both satisfy $\hat{d}\leq 2n$.

\textit{Example 4} ($K=5$):
According to \textit{Corollary \ref{corollary2}}, an IC system using the proposed partial coordination is proper for $K=5$ and can attain the upper bound on DOF, but, since $\hat{d} = \lfloor 5 n / 2 \rfloor > 2n$, it cannot be achieved in one shot. However, the iterative IA \cite{jafar2} algorithm operating on the equivalent IC model, i.e., $\prod_{k=1}^{5}{(m \times 2n,d_k)}$ can achieve it using the IA principles in \eqref{eqn3:io_userk_IC} and \eqref{eqn3:sol_req2}. The proposed one-shot solution can implement this system with a number of DOF that is under its theoretical upper bound, i.e., as a \textit{rigid} system with a total number of degrees of freedom of at most $2n$. For instance, the distributed IA \cite{jafar2} achieves the equivalent IC system of $\prod_{k=1}^{5}{(m \times 2n,d_k)}$ with $m=n=2$ with $5$ streams, i.e., at $\hat{d}=5$, iteratively, while, the proposed one-shot algorithm implements it with $4$ streams, i.e., at $\hat{d}=4$. In one-shot implementation, these $4$ streams can be shared among 5 user by employing the time sharing method described in Theorem \ref{theorem1} such that over the span of 5 time slots each user $k$ operates at $d_k = 4/5$ degrees of freedom. Note that within each one of the allocated 5 time slots, only 4 users are allowed to transmit while one user remains silent. This scenario for one-shot implementation can be considered as a special case for the $K=5$ system as it is effectively solved by modeling it as a $K=4$ system per time slot. As another example, the distributed IA \cite{jafar2} achieves the equivalent IC system of $\prod_{k=1}^{5}{(m \times 2n,d_k)}$ with $m=n=3$ with $7$ streams, i.e., at $\hat{d}=7$, through an iterative process, whereas, the one-shot algorithm implements it as a \textit{rigid} system with $6$ streams, i.e., $\hat{d}=6$,.
\\ 
These discussions and examples are confirmed through Monte-Carlo simulation results in section \ref{sec_sim}. 
\subsection{Overhead of CSI exchange in the backhaul}\label{sec:backhaul}
\begin{figure}[!htb]
\begin{center}
\includegraphics[width=.50\textwidth]{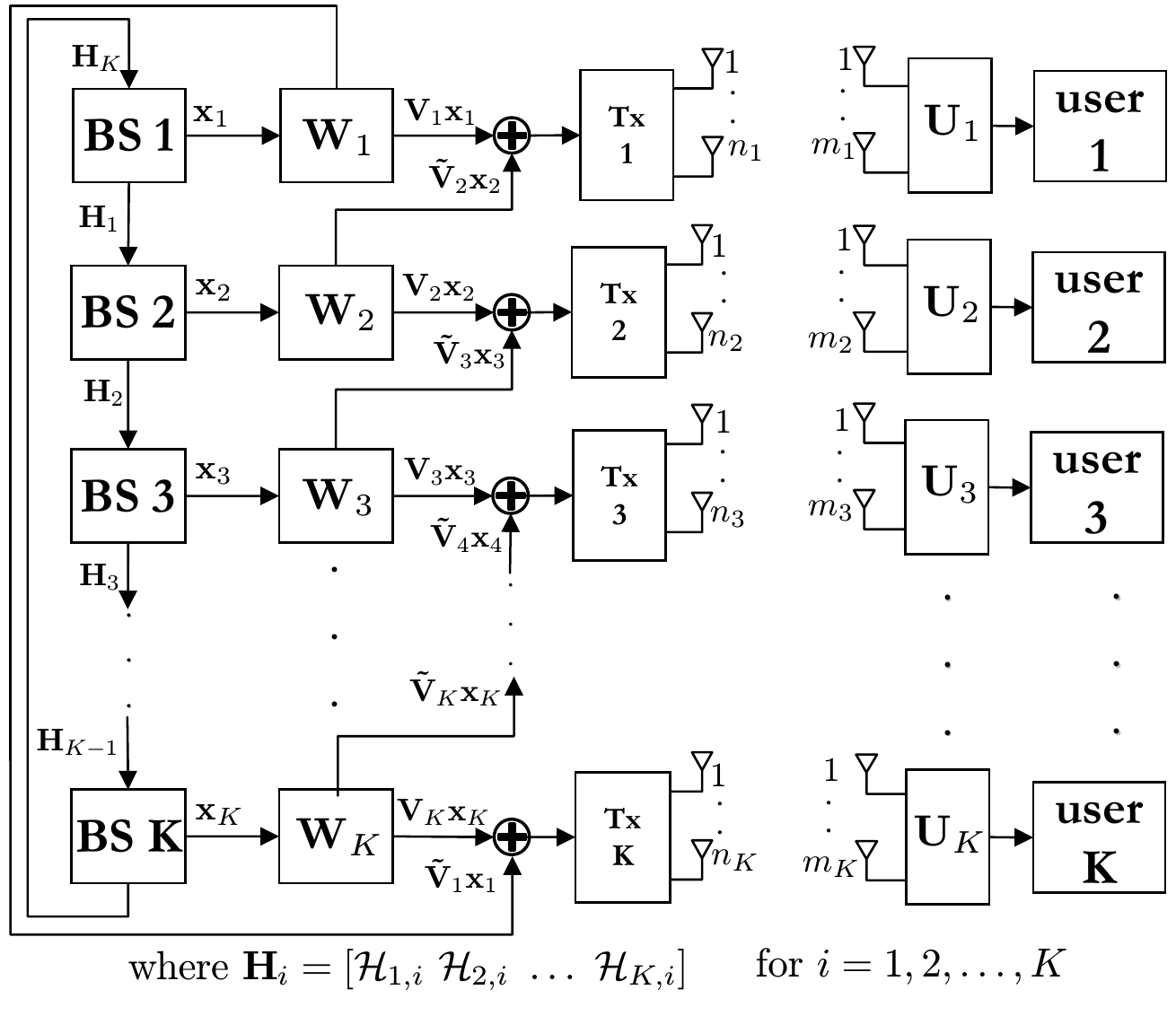}
\caption{\em System structure of the proposed partial coordination model in multicell downlink}
\label{fig:sysmodel}
\end{center}
\end{figure}

In practice, the backhaul network is built using dedicated point-to-point links (which can either be wired, e.g., fibre optic, or wireless, e.g., microwave) connecting all the BSs in a line or a ring topology. The implementation details of these topologies in wireless backhaul are discussed in chapter 6 of \cite{backhaul_book} The idea of using the line or the ring topology to model the multicell system was introduced by Wyner \cite{wyner}. This idea, commonly referred as Wyner linear or Wyner circular model, has been studied extensively in the context of joint multicell processing \cite{somekh1},\cite{somekh2}. In the following, we examine the data rate required for the exchange of CSI in the backhaul in the proposed coordination model.
\\
Let $\Delta t$ be the overall permissible latency to exchange CSI in the backhaul. Let $R$ be the required data rate to exchange the CSI of a single MIMO link, i.e., $\mathcal{H}_{i,j}$, $ \forall \ i,j$, between two coordinating BSs during $\Delta t$. Hence, in the ring model, a data rate of $KR$ is required in each and every BS-to-BS link to exchange the multiuser CSI, i.e., $\mathbf{H}_{i} = [ \mathcal{H}_{1,i} \ \ldots \ \mathcal{H}_{K,i}]$, $i=1,2,\ldots,K$, in Fig. \ref{fig:sysmodel}, within $\Delta t$. Whereas, in the line model, the inter-BS data rate of $2KR$ is required due to the broken link between the first and the $K^{th}$ coordinating BSs and the fact that the CSI of the $K^{th}$ BS, i.e., $\mathbf{H}_K$, should travel through all other links to reach the first BS. Therefore, the data rate requirement for CSI exchange in the backhaul grows linearly with $K$ in the proposed coordination structure in Fig. \ref{fig:sysmodel}. Whereas, in a system with fully coordinating BSs, e.g., \cite{zhang2004} - \cite{das}, it can be easily verified in a similar way that the data rate requirement for overall CSI exchange among all BSs within the same allowed latency interval of $\Delta t$ requires a data rate of $K^2R$ and $K(K-1)R$ in the line and the ring based backhaul models, respectively.
%
%
%
%
\section{Simulation Results}\label{sec_sim}
\begin{figure}[!htb]
\begin{center}
\includegraphics[width=.50\textwidth]{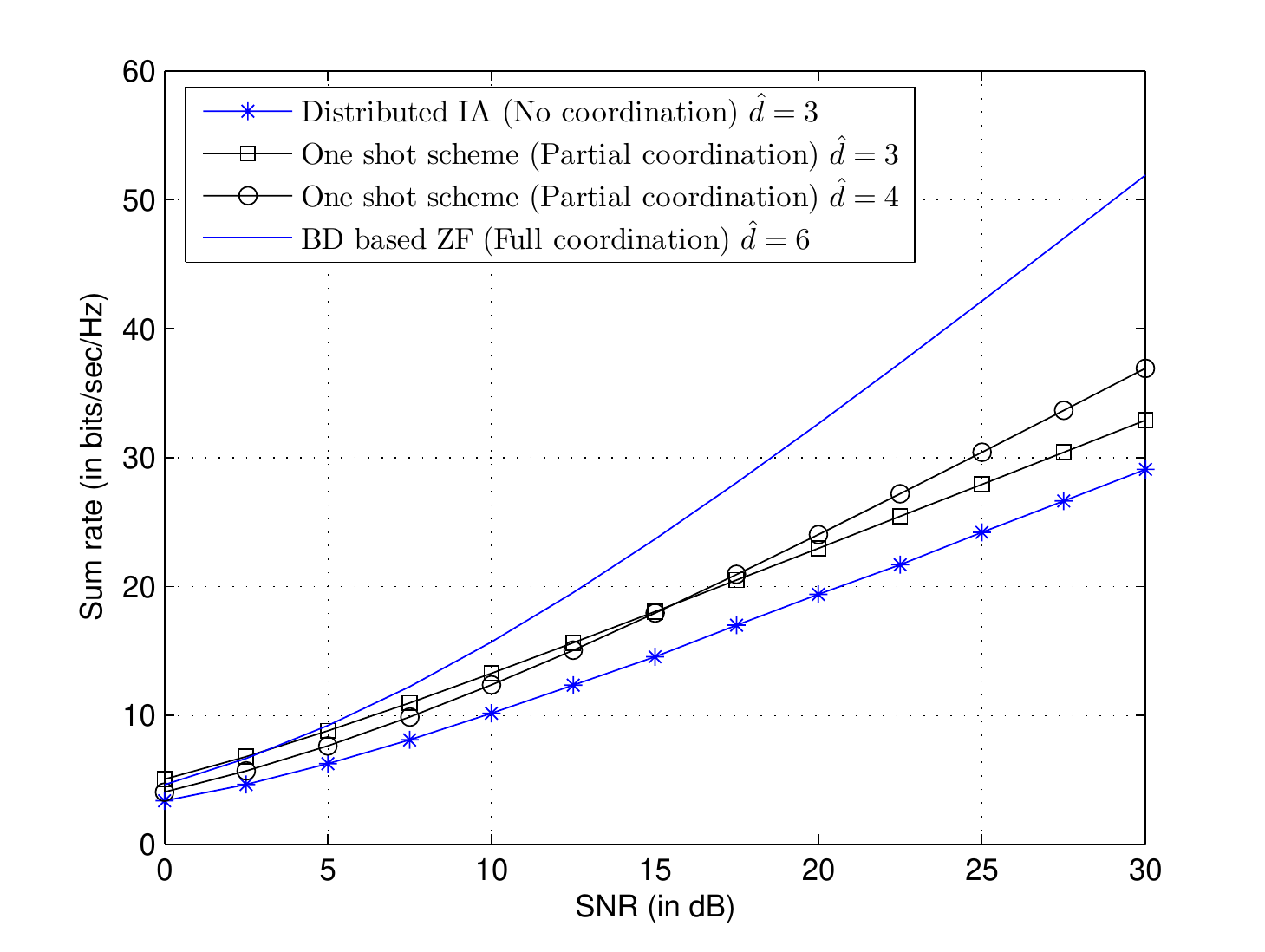}
\caption{\em Sum rate comparison of the partially coordinated one shot algorithm with distributed IA \cite{jafar2} (i.e., no coordination) and BD based ZF \cite{qhspence2} (with fully coordinating precoders), for $K=3$, with $m=n=2$.}
\label{fig:K3_2x2}
\end{center}
\end{figure}

\begin{figure}[!htb]
\begin{center}
\includegraphics[width=.50\textwidth]{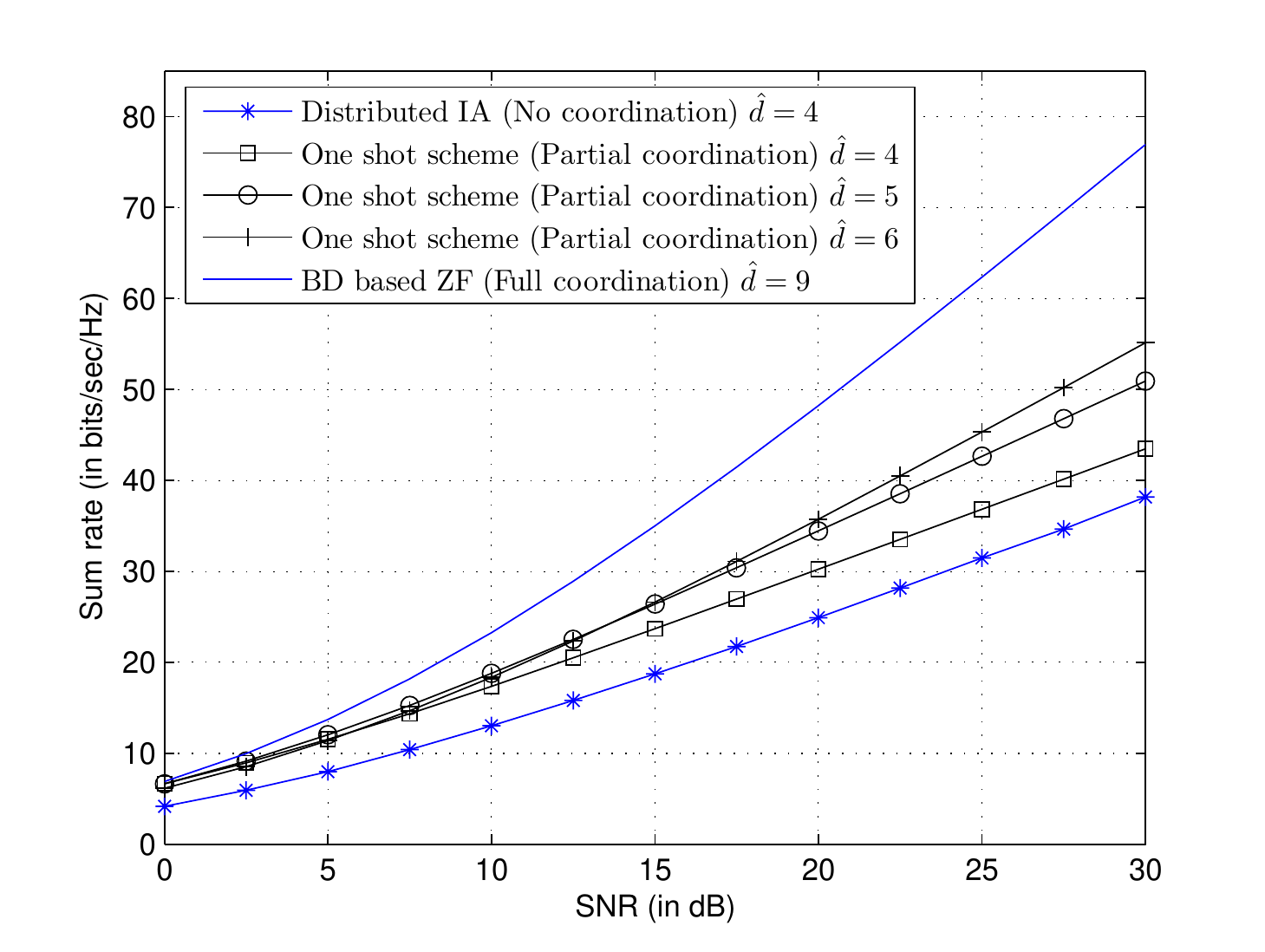}
\caption{\em Sum rate comparison of the partially coordinated one shot algorithm with distributed IA \cite{jafar2} (i.e., no coordination) and BD based ZF \cite{qhspence2} (with fully coordinating precoders), for $K=3$, with $m=n=3$.}
\label{fig:K3_3x3}
\end{center}
\end{figure}

\begin{figure}[!htb]
\begin{center}
\includegraphics[width=.50\textwidth]{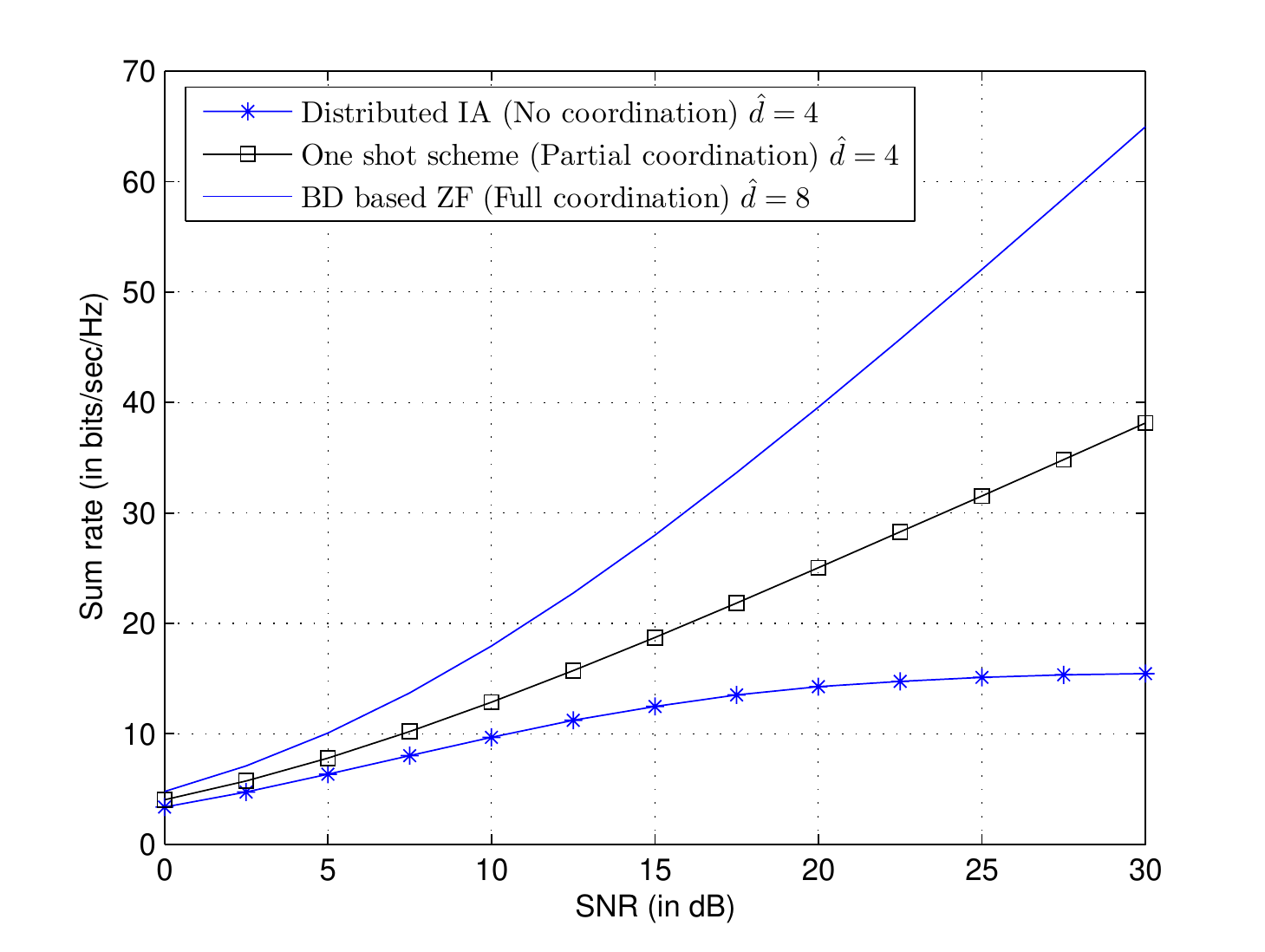}
\caption{\em Sum rate comparison of the partially coordinated one shot algorithm with distributed IA \cite{jafar2} (i.e., no coordination) and BD based ZF \cite{qhspence2} (with fully coordinating precoders), for $K=4$, with $m=n=2$.}
\label{fig:K4_2x2}
\end{center}
\end{figure}

\begin{figure}[!htb]
\begin{center}
\includegraphics[width=.50\textwidth]{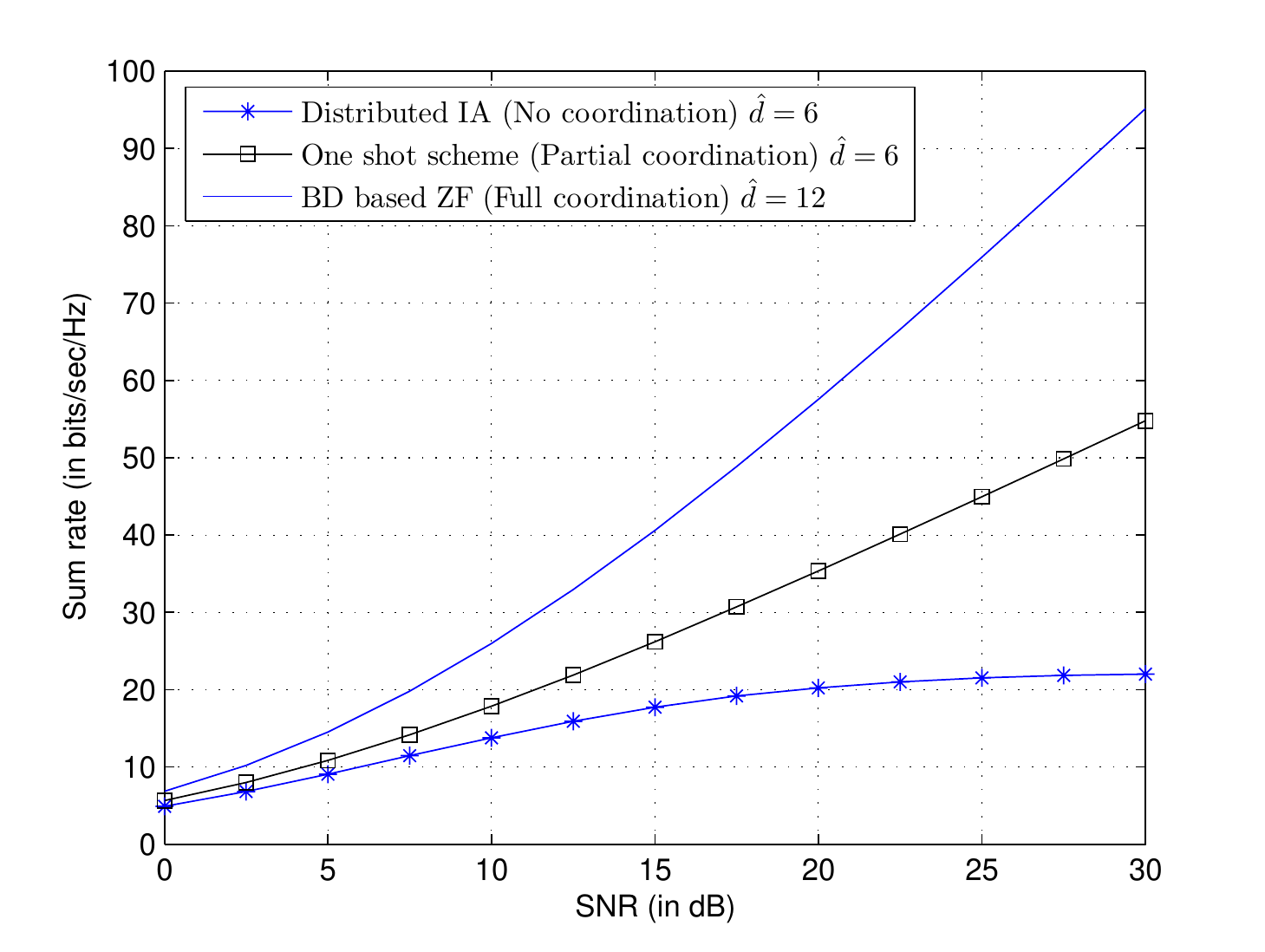}
\caption{\em Sum rate comparison of the partially coordinated one shot algorithm with distributed IA \cite{jafar2} (i.e., no coordination) and BD based ZF \cite{qhspence2} (with fully coordinating precoders), for $K=4$, with $m=n=3$.}
\label{fig:K4_3x3}
\end{center}
\end{figure}

\begin{figure}[!htb]
\begin{center}
\includegraphics[width=.50\textwidth]{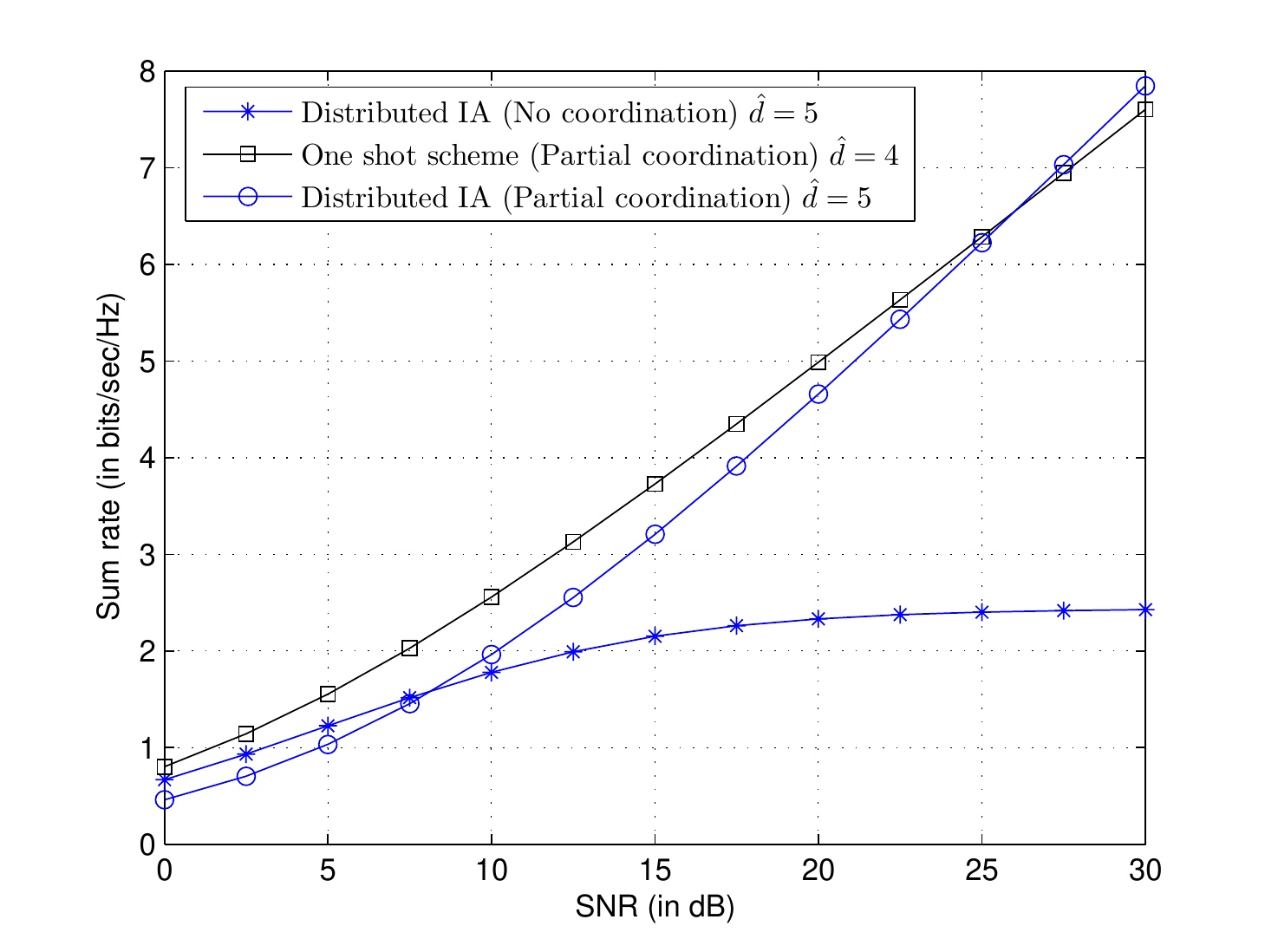}
\caption{\em Sum rate comparison of the partially coordinated one shot algorithm with distributed IA \cite{jafar2} in both partially coordinated and uncoordinated scenario, for $K=5$, with with $m=n=2$.}
\label{fig:K5_2x2}
\end{center}
\end{figure}

\begin{figure}[!htb]
\begin{center}
\includegraphics[width=.50\textwidth]{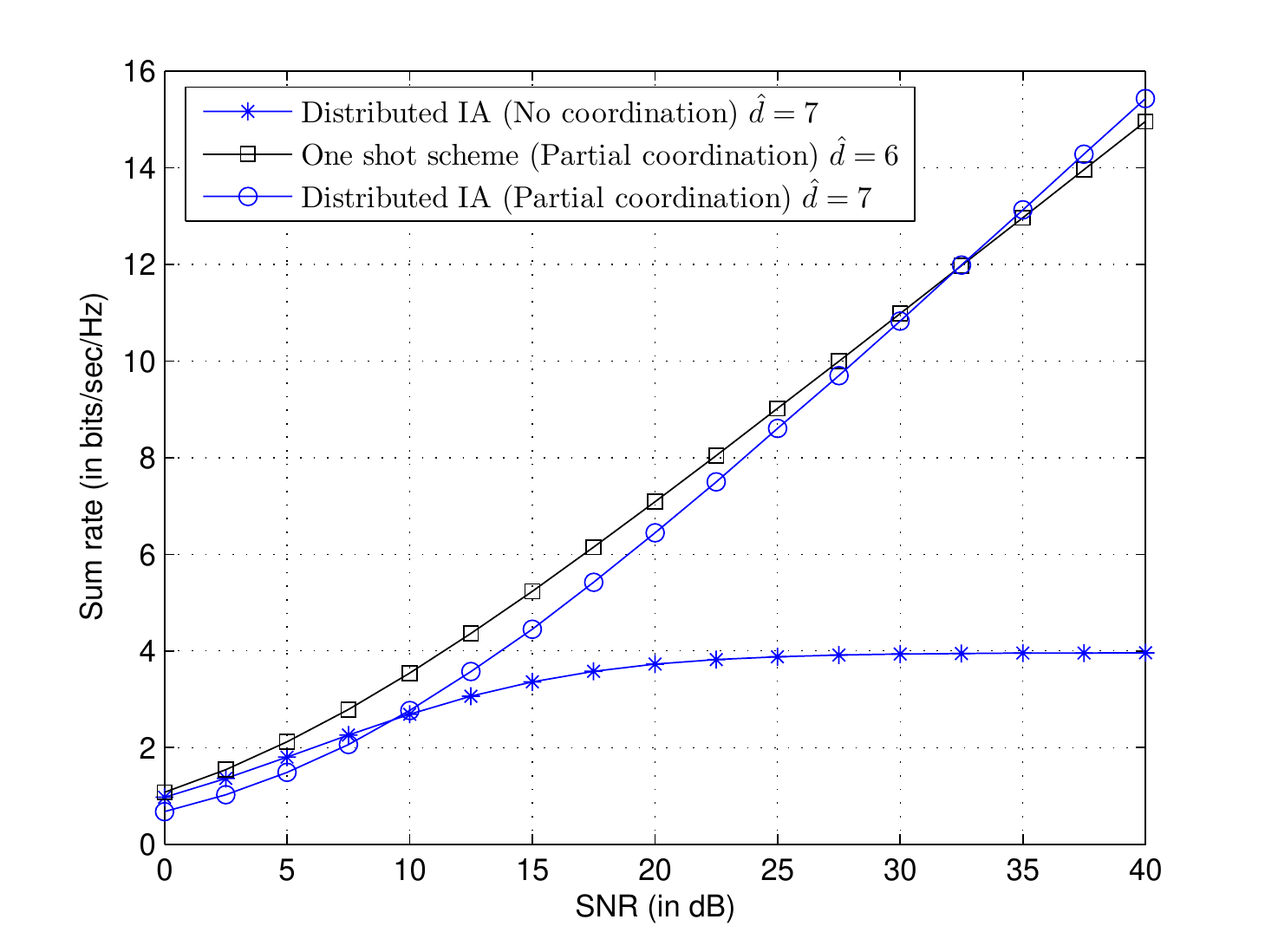}
\caption{\em Sum rate comparison of the partially coordinated one shot algorithm with distributed IA \cite{jafar2} in both partially coordinated and uncoordinated scenario, for $K=5$, with $m=n=3$.}
\label{fig:K5_3x3}
\end{center}
\end{figure}

\begin{figure}[!htb]
\begin{center}
\includegraphics[width=.50\textwidth]{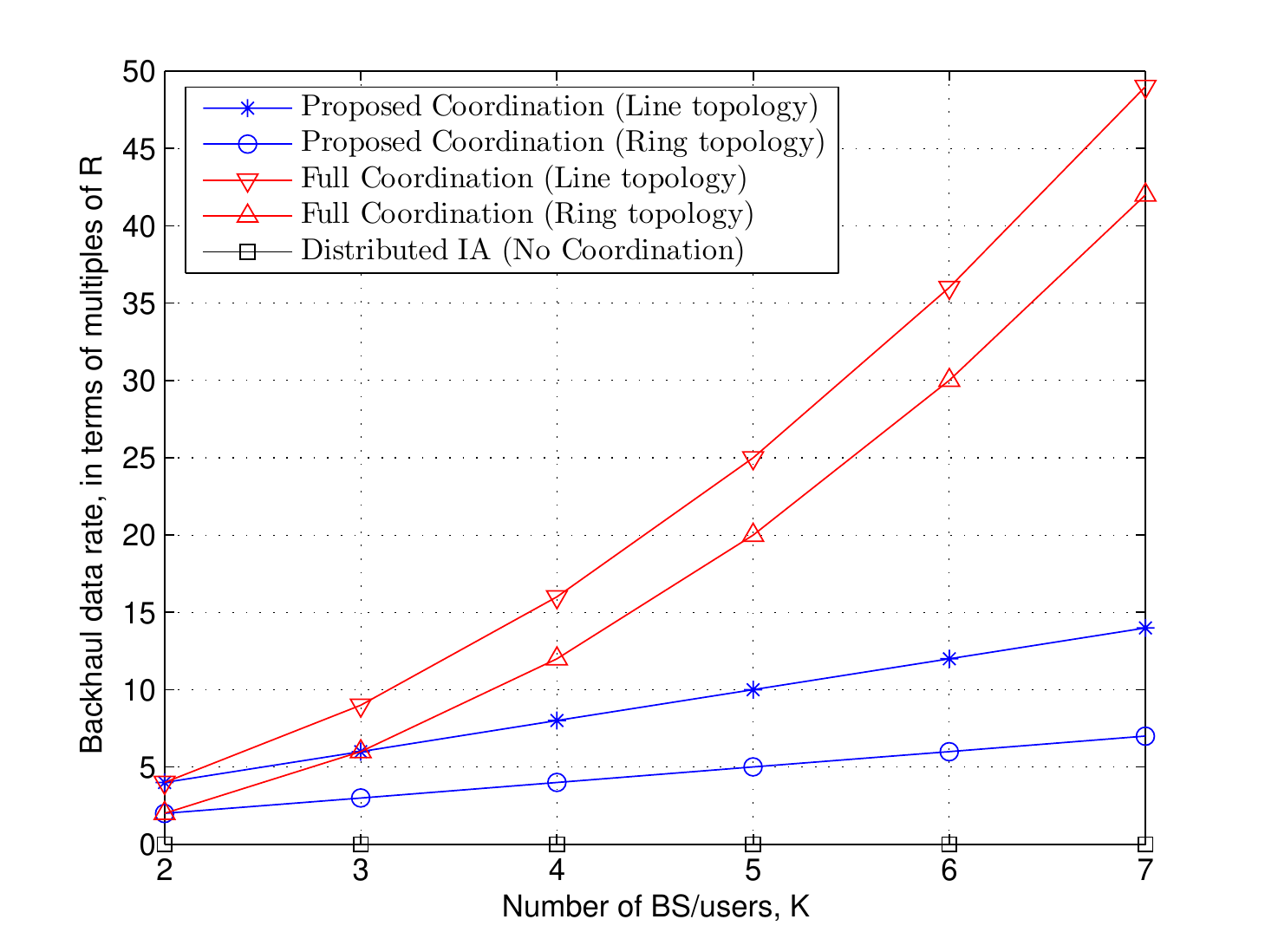}
\caption{\em Comparison of the data rate required in the backhaul for the exchange of CSI in the proposed partial coordination model with full coordination model and Distributed IA, i.e., non-coordinating scheme, for $K=2$ to $K=7$.}
\label{fig:backhaul}
\end{center}
\end{figure}

In this section, we report our simulation results on the achievable sum rate of the proposed partial coordination scheme and compare them with those of generic IC model, employing distributed IA algorithm \cite{jafar2}, and full coordination, i.e., block-diagonalization (BD) based zero-forcing (ZF) scheme \cite{qhspence2}. Transmitting BSs use the same transmit power and each BS equally distributes its power across its independent data streams. Figs. \ref{fig:K3_2x2}-\ref{fig:K5_3x3} compare the achievable sum rate of different schemes for $K=3$, $K=4$ and $K=5$, where $2 \times 2$ and $3 \times 3$ MIMO links are established between a BS and a single user in each cell. Results confirm that the BD based ZF in a full coordinated system, where separate base-stations form a giant single BS, achieves the highest sum rates amongst the other schemes. These results also confirm that the proposed one shot IA outperforms the distributed IA in terms of sum rate performance. It should be noted that degradation of performance of the distributed IA is because of the lack of coordination among the precoders. 
Furthermore, we compare the flexible and rigid systems for $K=3$, discussed as \textit{Example 2} in section \ref{subsec:gains}. In Fig. \ref{fig:K3_2x2}, i.e., $K=3$ with $2 \times 2$ MIMO links, the flexible and rigid systems operate at $\hat{d}=3$ and $\hat{d}=4$, respectively. Similarly, in Fig. \ref{fig:K3_3x3}, i.e., $K=3$ with $3 \times 3$ MIMO links, we have two flexible systems, with $\hat{d}=4$ and $\hat{d}=5$, and the rigid one operates at $\hat{d}=6$. Results confirm that systems with higher $\hat{d}$ offer higher sum rate at high SNR as the slope of their sum rate curves grow faster at high SNR, as shown in Figs. \ref{fig:K3_2x2} and \ref{fig:K3_3x3}. However, we observe that at lower SNR the flexible system, with lower DOF, outperforms the rigid one. This is due to the fact that in the flexible system the geometric mean of the individual SNRs is maximized through the selection of the transmit beamformers, whereas, in the rigid one the transmit beamformers are fixed. In Fig. \ref{fig:K3_2x2}, the rigid system, i.e., $\hat{d}=4$, offers higher sum rate compared to the flexible one, with $\hat{d}=3$, at SNRs higher than 15.5 dB. Similarly, in Fig. \ref{fig:K3_3x3}, the rigid system, i.e., $\hat{d}=6$, outperforms the flexible systems of $\hat{d}=4$ and $\hat{d}=5$ at SNRs higher than 6 dB and 13.5 dB, respectively. 
In Figs. \ref{fig:K4_2x2}-\ref{fig:K5_3x3}, the sum rate curves of distributed IA \cite{jafar2} in $K=4$ and $K=5$ with no coordination at the precoders saturate at higher SNRs due to leakage of interference into the interference free dimensions carrying the information streams. This confirms the result from \cite{jafar3} as these improper systems at $K=4$ and $K=5$, defined by \textit{Corollary \ref{corollary1}}, are unable to attain a feasible IA solution. 
In Figs. \ref{fig:K4_2x2} and \ref{fig:K4_3x3}, i.e., $K=4$ system with $2 \times 2$ and $3 \times 3$ MIMO links, the sum rate curves of the proposed one shot partially coordinated IA operating the information theoretic DOF upper bound, i.e., at $\hat{d}=4$ and $\hat{d}=6$, respectively, grows linearly with increasing SNR. This confirms our discussion in \textit{Example 3} as the proposed partially coordinated scheme solves $K=4$ at the information theoretic DOF upper bound in one shot.
Following the discussion from \textit{Example 4}, in Fig. \ref{fig:K5_2x2}, i.e., $K=5$ with $2 \times 2$ MIMO links, the distributed IA \cite{jafar2} achieves the information theoretic DOF upper bound at $\hat{d}=5$, i.e., $d_k=1$, $\forall \ k$, in the partially coordinated system whereas, our one shot solution achieves $\hat{d}=4$, i.e., $d_k=4/5$, $\forall \ k$, as described in Theorem \ref{theorem1}. Similarly, in Fig. \ref{fig:K5_3x3}, i.e., $K=5$ system with $3 \times 3$ MIMO links, the distributed IA \cite{jafar2} solves the DOF upper limit at $\hat{d}=7$, while the proposed one shot solution achieves $\hat{d}=6$. In Figs. \ref{fig:K5_2x2} and \ref{fig:K5_3x3}, the iterative IA, operating at $\hat{d}=5$ and $\hat{d}=7$, respectively, offers higher sum rate at high SNRs. However, we observe that at lower SNRs the proposed one shot IA operating at lower DOF, i.e., at $\hat{d}=4$ and $\hat{d}=6$, respectively, outperforms iterative IA. This improvement at lower SNRs can be intuitively explained as follows. In one shot IA, the receive beamformers in the original network (i.e., \textit{Step 1}) are designed to maximize the received signal power, whereas in distributed IA \cite{jafar2}, the beamformers are chosen to minimize interference leakage in both the original and the reciprocal networks. In Fig. \ref{fig:K5_2x2}, the partially coordinated distributed IA with $\hat{d}=5$ offers higher sum rate compared to one shot solution with $\hat{d}=4$ at SNRs higher than 26 dB. Similarly, in Fig. \ref{fig:K5_3x3}, the distributed IA with $\hat{d}=7$ outperforms the one shot algorithm with $\hat{d}=6$ at SNRs higher than 32 dB. 
Note that for $K=5$, the fully coordinated system offers very high multiplexing gain, i.e., $\hat{d}=10$ with $2 \times 2$ MIMO links and $\hat{d}=15$ with $3 \times 3$ MIMO links. Hence, to obtain a better resolution for the sum rate curves of the partially coordinated scheme, the sum rate curves for the fully coordinated BD based ZF are omitted in Figs. \ref{fig:K5_2x2} and \ref{fig:K5_3x3}.
In Fig. \ref{fig:backhaul}, we compare the data rate required for exchanging the CSI overhead in the backhaul in the proposed partially coordinated precoding scheme with non-coordinating interference channel model and the full coordination model \cite{zhang2004}-\cite{das}, as discussed in section \ref{sec:backhaul}. 
The data rates are expressed in terms of multiples of $R$, where $R$ is the data rate required to send the single MIMO link CSI through the backhaul within the allowed latency interval $\Delta t$. 
The full coordination model requires exponential increase in backhaul data rate, whereas, the backhaul data rate of the partially coordinated IA increases linearly with $K$.
%
%
\section{Conclusions}
In this paper, we introduced a partially coordinated transmit precoding scheme for downlink multicell MIMO systems. We transformed the system model of our scheme to an equivalent interference channel problem and compared its feasibility with standard interference channel model. We established a sufficient condition for obtaining a proper system, in terms of the maximum number of users supported at the information theoretic upper bound on DOF, in the proposed partially coordinated IA and the generic IA. Furthermore, we introduced a one shot IA algorithm with partial coordination and analyzed the maximum achievable degrees of freedom in the system. Simulation results confirmed that, in comparison with the generic IA, the proposed partially coordinated IA offers higher throughput at practical SNR levels and avoids an uncontrolled number of back and forth iterations between the original and the reciprocal networks.  
%
%
%
%


%
\newpage

%
%
%
%
%
%

\end{document}